\documentclass{ws-procs9x6-cpt25}

\usepackage{slashed}
\usepackage{natbib}
\usepackage{bbm}

\begin{document}

\newcommand{\refeq}[1]{(\ref{#1})}
\def\etal {{\it et al.}}

\newcommand{\CLsing}{{}^{1}C_L}
\newcommand{\CRsing}{{}^{1}C_R}
\newcommand{\CLtrip}{{}^{3}C_L}
\newcommand{\CRtrip}{{}^{3}C_R}
\newcommand{\CLRsing}{{}^{1}C_{L/R}}
\newcommand{\CLRtrip}{{}^{3}C_{L/R}}
\newcommand{\tCLtrip}{{}^{3}\widetilde{C}_L}
\newcommand{\tCRtrip}{{}^{3}\widetilde{C}_R}
\newcommand{\tCLRtrip}{{}^{3}\widetilde{C}_{L/R}}

\title{Hadronic Lorentz violation in chiral perturbation theory}

\author{M.R.\ Schindler$^1$}

\address{$^1$Department of Physics and Astronomy, University of South Carolina,\\
Columbia, SC, 29208, USA}

\begin{abstract}
Lorentz violation in hadronic systems is related to Lorentz-violating operators of quarks and gluons. 
Due to the nonperturbative nature of quantum chromodynamics (QCD) at low energies, establishing these relationships is complex. Chiral perturbation theory (ChPT) is an effective theory that provides one method of connecting quark- and gluon-level operators to those at the hadronic level, which can be used to calculate hadronic observables. 
\end{abstract}

\bodymatter

\section{Introduction}
Lorentz violation in the strongly-interacting sector of the Standard-Model Extension (SME)\cite{Colladay:1996iz,Colladay:1998fq} is formulated in terms of quark and gluon degrees of freedom. 
On the other hand, experiments on strongly-interacting particles are performed on color-neutral hadrons.
Because quantum chromodynamics (QCD) is nonperturbative at low energies, relating the properties and interactions of hadrons to those of the underlying quark and gluon degrees of freedom is challenging.
This challenge already arises in the Lorentz-invariant conventional Standard Model context.
Chiral perturbation theory (ChPT) \cite{Weinberg:1978kz,Gasser:1983yg,Gasser:1984gg} is an effective field theory of QCD, which is defined in terms of mesons and baryons. 
It is based on the (approximate) symmetries of QCD and is valid at energies well below $1\,\text{GeV}$. For a pedagogical introduction to ChPT see, e.g., Ref.~\refcite{Scherer:2012xha}.
ChPT has also been extended to incorporate the effects of Lorentz-violating (LV) quark and gluon operators in terms of hadronic operators.
In the following I describe the basic principles behind this approach and give an example of how ChPT can help in setting new bounds on LV parameters.
 
\section{Basics of chiral perturbation theory}

\subsection{Chiral symmetry}
Chiral perturbation theory is based on the symmetries of QCD. 
In addition to $\text{SU}(3)_c$ color gauge symmetry and the discrete symmetries of parity (P), time reversal (T), and charge conjugation (C),\footnote{The QCD $\theta$ term is ignored in the following.} QCD exhibits an approximate chiral symmetry originating in the light quark masses.
When considering processes at energies well below $1\, \text{GeV}$, the heavy quarks $c$, $b$, and $t$ do not have to be included explicitly. 
The QCD Lagrangian for the light quark flavors is given by
\begin{align}\label{QCDLag}
    \mathcal{L}_{QCD} = \sum_{l} \bar{q}_l (i\slashed{D}-m_l) q_l -\frac{1}{2}\text{Tr}(\mathcal{G}^{\mu\nu}\mathcal{G}_{\mu\nu}) ,
\end{align}
where the index $l$ denotes the light quark flavors, $q_l$ are the light quark fields, and $\mathcal{G}_{\mu\nu}$ denotes the gluon field strength tensor. The following discussion is mainly restricted to up and down quarks ($l=u,d$), but the inclusion of the strange quark ($l=u,d,s$) is also possible.
Because the light quark masses $m_l$ are much smaller than typical hadronic scales, setting them to zero can serve as a useful theoretical starting point. 
This limit of $m_l \to 0$ is referred to as the chiral limit.
It is convenient to collect the light quark flavors into a two-component  multiplet $Q = (q_u, q_d)^T$ (or a three-component multiplet $(q_u, q_d,q_s)^T$ if the strange quark is included) and to introduce right- and left-handed quark fields,
\begin{align}
    Q_R = \frac{1}{2}(\mathbbm{1}+\gamma_5)Q,\quad Q_L = \frac{1}{2}(\mathbbm{1}-\gamma_5)Q,\quad  \gamma_5=i\gamma^0\gamma^1\gamma^2\gamma^3.
\end{align}
The QCD Lagrangian in the chiral limit then takes the form
\begin{align}\label{QCDLagchiral}
    \mathcal{L}^0_{QCD} =  \bar{Q}_{L} i\slashed{D} Q_{L} +\bar{Q}_{R} i\slashed{D} Q_{R} -\frac{1}{2}\text{Tr}(\mathcal{G}^{\mu\nu}\mathcal{G}_{\mu\nu}) .
\end{align}
At the classical level, this Lagrangian is invariant under independent global U(2) transformations of the left- and right-handed fields,
\begin{align}\label{chiraltrans}
    Q_L\mapsto U_L Q_L,\quad Q_R \mapsto U_R Q_R,
\end{align}
where $U_{L/R} \in \text{U(2)}$, i.e., $\mathcal{L}^0_{QCD}$ possesses a classical $\text{U(2)}_L\times \text{U(2)}_R$ symmetry.
Because of the axial anomaly, at the quantum level this symmetry reduces to an $\text{SU(2)}_L\times \text{SU(2)}_R \times \text{U(1)}_V$ symmetry, where the $\text{U(1)}_V$ symmetry is related to baryon number conservation and the invariance under $\text{SU(2)}_L\times \text{SU(2)}_R$ is referred to as chiral symmetry.
One expected consequence of this chiral symmetry is the existence of degenerate hadron multiplets of opposite parity. 
The absence of this parity doubling combined with the observation that pions are much lighter than all other hadrons leads to the assumption that chiral symmetry is spontaneously broken to an $\text{SU(2)}_V$ symmetry and that the pions are the Goldstone bosons of this spontaneous symmetry breaking.

In addition to the assumed spontaneous breaking of chiral symmetry, the finite light quark masses break chiral symmetry explicitly. 
The quark mass contribution to the Lagrangian of Eq.~\refeq{QCDLag} takes the form
\begin{align}\label{QCDMassLag}
    \mathcal{L}_M = -(\bar{Q}_R\mathcal{M} Q_L + \bar{Q}_L\mathcal{M}^\dagger Q_R),
\end{align}
where 
\begin{align}\label{MassMatrix}
    \mathcal{M} = \begin{pmatrix}
        m_u & 0\\ 0 & m_d
    \end{pmatrix}
\end{align}
is the mass matrix for the light quarks. 
Because $\mathcal{L}_M$ couples left- to right-handed fields, it is not invariant under chiral transformations.
As a consequence of this explicit chiral symmetry breaking due to the quark masses,  the pions as the Goldstone bosons of spontaneous chiral symmetry breaking are not massless. 

\subsection{ChPT for mesons}
ChPT as the EFT of QCD at low energies incorporates the symmetries of QCD and their spontaneous and explicit breaking, but is formulated in terms of pions and nucleons instead of quarks and gluons.
Chiral symmetry is realized nonlinearly on the pions,\cite{Weinberg:1978kz,Coleman:1969sm} which are collected in the SU(2) matrix
\begin{align}\label{Udef}
    U(x) = \exp\left( i \frac{\phi(x)}{F} \right),
\end{align}
with
\begin{align}
    \phi(x) = \sum_{a=1}^3 \phi_a \tau_a =
    \begin{pmatrix}
         \pi^0 & \sqrt{2} \pi^+ \\
         \sqrt{2} \pi^- & \pi^0
    \end{pmatrix},
\end{align}
and the quantity $F$ is identified with the pion decay constant in the chiral limit.
Under chiral transformations $(L,R) \in \text{SU(2)}_L \times \text{SU(2)}_R $, the matrix of Eq.~\refeq{Udef} transforms as
\begin{align}
    U(x) \mapsto R^\dagger U(x) L.
\end{align}
In the chiral limit, i.e., for zero quark masses, the ChPT Lagrangian is formulated in terms of the fields $U$ and its derivatives. 
As long as one restricts the discussion to momenta much less than the scale $4\pi F \approx 1\,\text{GeV}$, the effective Lagrangian can be expanded in the number of derivatives acting on $U$. 
The term with the lowest number of derivatives (aside from a constant) that is invariant under the assumed symmetries is given by
\begin{align}
    \frac{F^2}{4}\text{Tr}\left[ \partial_\mu U (\partial^\mu U)^\dagger \right].
\end{align}
Expanding the matrix $U$ in the pion fields leads not only to a canonical kinetic term for massless pions, but also to interaction terms with four and more pion fields. 
That all of these terms are related to a single low-energy coefficient (LEC) $F$ is a consequence of chiral symmetry.

The inclusion of the quark mass term in the ChPT Lagrangian serves as the prototype for including LV effects. 
While the Lagrangian of Eq.~\refeq{QCDMassLag} is not invariant under chiral transformations, it \emph{would} be invariant \emph{if}
\begin{equation}
    \mathcal{M} \mapsto R\mathcal{M} L^\dagger.
\end{equation}
The mass matrix $\mathcal{M}$, being a constant matrix, does not actually have this transformation behavior.
However, by treating it as a spurion field with this behavior and constructing chirally invariant Lagrangians with it, the pattern of chiral symmetry breaking is mapped from the QCD term to the chiral Lagrangian.
Imposing Lorentz invariance and invariance under discrete symmetries, the term with a single power of $\mathcal{M}$ is given by
\begin{align}
    \frac{F^2 B^2}{2}\text{Tr}\left[ \mathcal{M} U^\dagger + U \mathcal{M}^\dagger  \right].
\end{align}
The new LEC $B$ is related to the scalar quark condensate and relates the quark masses to the lowest-order ChPT result for the pion masses, which assuming isospin symmetry $m_u = m_d = \hat{m}$ takes the form $M_{\pi,2}^2 = 2 B \hat{m}$.

The complete leading-order (LO) ChPT Lagrangian for pions is 
\begin{align}
    \mathcal{L}_2 = \frac{F^2}{4}\text{Tr}\left[ \partial_\mu U (\partial^\mu U)^\dagger \right] +\frac{F^2 B^2}{2}\text{Tr}\left[ \mathcal{M} U^\dagger + U \mathcal{M}^\dagger  \right].
\end{align}
ChPT is based on a simultaneous expansion of observables in derivatives and quark masses, divided by the breakdown scale $\Lambda_b$ of the EFT. 
It is also possible to consider the coupling to external fields, such as electromagnetic or weak gauge fields. For this purpose it is convenient to promote the \emph{global} chiral symmetry to a \emph{local} symmetry.\cite{Gasser:1983yg,Gasser:1984gg}

\subsection{ChPT for baryons}
Nucleons (and strange baryons) can also be treated in ChPT.\cite{Gasser:1987rb}
The proton (p) and neutron (n)  are collected into a doublet $\psi = (p,n)^T$, which transforms under chiral transformations as
\begin{align}
    \psi \mapsto K(L,R,U) \psi,
\end{align}
where 
\begin{align}
    K(L,R,U) = \sqrt{R U L^\dagger}^{-1} R \sqrt{U}
\end{align}
is an SU(2) matrix that not only depends on the left- and right-handed transformations $L$ and $R$, but also on the pion fields through the matrix $U$.
The lowest-order nucleonic Lagrangian is given by
\begin{align}\label{LOpiNLag}
    \mathcal{L}_{\pi N}^{(1)} = \bar{\psi} \left(i\slashed{D}-m+\frac{g_A}{2} \gamma^\mu \gamma_5 u_\mu \right)\psi ,
\end{align}
with the covariant derivative on the nucleon fields defined as
\begin{align}
    D_\mu \psi = \left(\partial_\mu - \Gamma_\mu \right)\psi,\quad \Gamma_\mu = \frac{1}{2}(u^\dagger \partial_\mu u + u \partial_\mu u^\dagger),
\end{align}
and the chiral vielbein
\begin{align}
    u_\mu = i \left( u^\dagger \partial_\mu u - u \partial_\mu u^\dagger  \right).
\end{align}
The matrix $u$ is the square root of the pion matrix $U$, i.e., $u^2 = U$.
The Lagrangian of Eq.~\refeq{LOpiNLag} contains two LECs: $m$, the nucleon mass in the chiral limit, and $g_A$, the nucleon axial-vector coupling constant, also taken in the chiral limit.
Expanding $u$ in terms of pion fields leads to a tower of pion-nucleon interaction terms.
Unlike for pions, the nucleon mass in the chiral limit does \emph{not} vanish in the chiral limit, which makes renormalization and power counting for nucleons more complex than in the mesonic sector.

\section{Lorentz violation in ChPT}

The ChPT formalism has been applied to include the effects of LV operators from the minimal SME.\cite{Noordmans:2016pkr,Kamand:2016xhv,Noordmans:2017kji,Kamand:2017bzl,Altschul:2019beo}
The following presents a representative example taken from Ref.~\refcite{Kamand:2016xhv} of how LV operators can be incorporated in ChPT; a more detailed discussion and applications to other operators, as well as the coupling to external fields, is found in the literature.

Consider the CPT-even, dimension-4 quark operators with LV parameters $c_{\mu\nu}$.
After a naive restriction to two light flavors, the quark-level Lagrangian takes the form
\begin{align}
    \mathcal{L}^{\text{LV}}_\text{quark}= \frac{i}{2} \bar{Q}_{L} C_L^{\mu\nu} \gamma_{\mu} \overset{\leftrightarrow}{\mathcal{D}_\nu} Q_{L}
+ \frac{i}{2} \bar{Q}_{R} C_R^{\mu\nu}  \gamma_{\mu} \overset{\leftrightarrow}{\mathcal{D}_\nu} Q_{R},
\end{align}
where $\bar{Q}\overset{\leftrightarrow}{\mathcal{D}_\nu} Q$ is the covariant derivative acting on the quark fields and the LV parameters are collected in the matrices
\begin{align}
    C_{L/R}^{\mu\nu}=\begin{pmatrix}
c^{\mu\nu}_{u_{L/R}} & 0 \\
0 & c^{\mu\nu}_{d_{L/R}}
\end{pmatrix}.
\end{align}
In addition to violating Lorentz invariance, $\mathcal{L}^{\text{LV}}_\text{quark}$ is \emph{not} invariant under chiral transformations,
$$
Q_L \mapsto L Q_L, \quad Q_R \mapsto R Q_R.
$$
Similar to the case of a finite quark-mass matrix, the constant matrices $C^{\mu\nu}_{L/R}$ are promoted to spurion fields: the quark-level LV Lagrangian is invariant if
the $C^{\mu\nu}_{L/R}$ are assumed to transform under chiral transformations as
\begin{align}
    C_L^{\mu\nu} \mapsto L C_L^{\mu\nu} L^\dagger, \quad C_R^{\mu\nu} \mapsto R C_R^{\mu\nu} R^\dagger.
\end{align}
It is also convenient to separate these matrices into their isoscalar and isovector components,
$$
    C_{L/R}^{\mu\nu} = \CLRsing^{\mu\nu} \mathbbm{1} +\CLRtrip^{\mu\nu} \tau_3.
$$

With this assumed transformation behavior, we can now construct chirally invariant Lagrangians in terms of pion and nucleon fields that map Lorentz violation from the quark to the hadronic level.
Assuming that the $C^{\mu\nu}_{L/R}$ are traceless and symmetric in the Lorentz indices, there is a single term at LO in the chiral power counting in the purely mesonic sector,
\begin{align}
    \mathcal{L}_{\pi}^{\text{LV,LO}} = 
\beta^{(1)}\frac{F^2}{4}\left({\CRsing}_{\mu\nu} +{\CLsing}_{\mu\nu}\right)
\text{Tr}[(\partial^{\mu}U)^{\dagger} \partial^{\nu}U] .
\end{align}
The parameter $\beta^{(1)}$ is a new LEC that incorporates short-distance physics. It is dimensionless and, based on naive dimensional analysis, is expected to be $\mathcal{O}(1)$.
A term containing $\CLRtrip^{\mu\nu}$ vanishes for symmetric $C^{\mu\nu}_{L/R}$.

The structure in the nucleonic sector is more complex. At LO, the Lagrangian is given by
\begin{align}
\begin{split}
    \mathcal{L}_{\pi N}^{\text{LV,LO}}= & \Big\{
\alpha^{(1)}\bar{\Psi}[(\tCRtrip^{\mu\nu} + \tCLtrip^{\mu\nu}) (\gamma_{\nu} i D_{\mu} + \gamma_{\mu} i D_{\nu})]\Psi \\
 & + \alpha^{(2)}\left({\CRsing^{\mu\nu}} + \CLsing^{\mu\nu}\right)\bar{\Psi}(\gamma_{\nu} i D_{\mu} + \gamma_{\mu}i D_{\nu})]\Psi  \\
 & + \alpha^{(3)}\bar{\Psi}[(\tCRtrip^{\mu\nu} - \tCLtrip^{\mu\nu}) (\gamma_{\nu}\gamma^{5} i D_{\mu} + \gamma_{\mu}\gamma^{5} i
 D_{\nu})]\Psi  \\
 & + \alpha^{(4)}\left(\CRsing^{\mu\nu} - \CLsing^{\mu\nu}\right) \bar{\Psi} (\gamma_{\nu}\gamma^{5} i D_{\mu} +
 \gamma_{\mu}\gamma^{5} i D_{\nu})\Psi\Big\} + \text{H.c.},
 \end{split}
\end{align}
where 
$$
\tCLtrip^{\mu\nu} = u \CLtrip^{\mu \nu} u^\dagger, \quad \tCRtrip^{\mu\nu} = u^\dagger \CRtrip^{\mu\nu} u.
$$
The four independent LECs $\alpha^{(i)}$ are again expected to be $\mathcal{O}(1)$.

The LECs $\beta^{(1)}$ and $\alpha^{(i)}$ are all related to the underlying quark-level LV parameters ${c_{u/d}^{\mu\nu}}_{L/R}$.
A determination of the effective hadronic LECs in terms of the quark-level parameters requires a nonperturbative QCD calculation, which in principle might be feasible using lattice QCD.
But even in the absence of such a calculation the ChPT expressions are still useful.
Expanding the matrices $U$ and $u$ in the number of pion fields gives the LV corrections to the pion and nucleon propagators as well as LV interaction vertices.
For the pions, the propagator corrections take the form 
\begin{align}
    \mathcal{L}_{\pi}^{\text{LV,LO},2\phi}=\frac{\beta^{(1)}}{4}(c^{\mu\nu}_{u_{L}}+c^{\mu\nu}_{d_{L}}+
c^{\mu\nu}_{u_{R}}+c^{\mu\nu}_{d_{R}})\partial_{\mu}\phi_{a}\partial_{\nu}\phi_{a}.
\end{align}
Comparison with the canonical form for a spin-0 particle,
\begin{align}
    \mathcal{L}_{\text{spin-0}}^\text{LV}=\frac{1}{2}k^{\mu\nu}\partial_{\mu}\phi_{a}\partial_{\nu}\phi_{a},
\end{align}
gives the ChPT result for the relationship between the meson- and quark-level parameters:
\begin{align}\label{kpi}
    k_{\pi}^{\mu\nu}=\frac{\beta^{(1)}}{2}
(c_{u_{L}}^{\mu\nu}+c_{u_{R}}^{\mu\nu}+c_{d_{L}}^{\mu\nu}+c_{d_{R}}^{\mu\nu}).
\end{align}
Analogous steps lead to the ChPT expressions relating the proton and neutron $c^{\mu\nu}_{p/n}$ coefficients to the quark-level parameters; e.g., for the proton
\begin{align}
    c_{p}^{\mu\nu}=\frac{1}{2}\left[\alpha^{(1)}+\alpha^{(2)}\right] (c_{u_{L}}^{\mu\nu}+c_{u_{R}}^{\mu\nu})
+\frac{1}{2}\left[-\alpha^{(1)}+\alpha^{(2)}\right](c_{d_{L}}^{\mu\nu}+c_{d_{R}}^{\mu\nu}).
\end{align}
Summing the expressions for the proton and neutron yields a result that is proportional to the same linear combination of quark-level parameters as the pion coefficient in Eq.~\refeq{kpi},
\begin{align}
    c_{p}^{\mu\nu}+c_{n}^{\mu\nu} \propto c_{u_{L}}^{\mu\nu}+c_{u_{R}}^{\mu\nu}+c_{d_{L}}^{\mu\nu}+c_{d_{R}}^{\mu\nu}.
\end{align}
Using the assumption that the ChPT LECs $\beta^{(1)}$ and $\alpha^{(i)}$ are all expected to be $\mathcal{O}(1)$, one can use existing bounds on the proton and neutron parameters to constrain the pion parameter $k_\pi^{\mu\nu}$. 
Even a conservative assumption about the relative sizes of $\beta^{(1)}$ and the $\alpha^{(i)}$ results in improvements on the pion parameters by several orders of magnitude.\cite{Kamand:2016xhv}

\section{Conclusions}

Chiral perturbation theory is the EFT of QCD for energies below $1\,\text{GeV}$, formulated in terms of color-neutral hadrons. 
It has been extended to include LV operators. 
ChPT establishes a connection between the LV operators of the SME, which are formulated in terms of quarks and gluons, and the hadronic-level operators that are used to analyze experiments.
The ChPT approach is based on the approximate chiral symmetry of QCD and the pattern of chiral symmetry breaking that the SME operators exhibit.
Chiral symmetry imposes relationships on operators with varying numbers of fields, e.g., relating corrections to the pion propagator to vertices with four and more pions.
ChPT thus provides a very useful tool for understanding how quark-level LV operators manifest themselves at the hadronic level, and its use has the promise to establish many new bounds on Lorentz violation in strongly-interacting particles.

\section*{Acknowledgments}
I thank my collaborators R.~Kamand and B.~Altschul.  
This material is based upon work supported by the U.S.~Department of Energy, Office of Science, Office of Nuclear Physics, under Award Numbers DE-SC0010300 and DE-SC0019647.


\end{document}